\documentclass[8.5pt,twoside,twocolumn]{article}
\usepackage[super,sort&compress,comma]{natbib} 
\usepackage{mhchem}
\usepackage{times,mathptmx}
\usepackage{sectsty}
\usepackage{balance} 

\usepackage{graphicx} 
\usepackage{lastpage}
\usepackage[format=plain,justification=raggedright,singlelinecheck=false,font=small,labelfont=bf,labelsep=space]{caption} 
\usepackage{fancyhdr}
\begin{document}

\thispagestyle{plain}
\fancypagestyle{plain}{
\renewcommand{\headrulewidth}{1pt}}
\renewcommand{\thefootnote}{\fnsymbol{footnote}}
\renewcommand\footnoterule{\vspace*{1pt}%
\hrule width 3.4in height 0.4pt \vspace*{5pt}} 
\setcounter{secnumdepth}{5}

\makeatletter 
\def\subsubsection{\@startsection{subsubsection}{3}{10pt}{-1.25ex plus -1ex minus -.1ex}{0ex plus 0ex}{\normalsize\bf}} 
\def\paragraph{\@startsection{paragraph}{4}{10pt}{-1.25ex plus -1ex minus -.1ex}{0ex plus 0ex}{\normalsize\textit}} 
\renewcommand\@biblabel[1]{#1}            
\renewcommand\@makefntext[1]%
{\noindent\makebox[0pt][r]{\@thefnmark\,}#1}
\makeatother 
\renewcommand{\figurename}{\small{Fig.}~}
\sectionfont{\large}
\subsectionfont{\normalsize} 

\fancyfoot{}
\fancyfoot[LO,RE]{\vspace{-7pt}}
\fancyfoot[CO]{\vspace{-7.2pt}\hspace{10cm}Cite this: Lab on a Chip, 2012, 3380-3386}
\fancyfoot[CE]{\vspace{-7.5pt}\hspace{-11cm}Cite this: Lab on a Chip, 2012, 3380-3386}
\fancyfoot[RO]{\footnotesize{\sffamily{~\textbar  \hspace{2pt}\thepage}}}
\fancyfoot[LE]{\footnotesize{\sffamily{\thepage~\textbar\hspace{9cm} }}}
\fancyhead{}
\renewcommand{\headrulewidth}{1pt} 
\renewcommand{\footrulewidth}{1pt}
\setlength{\arrayrulewidth}{1pt}
\setlength{\columnsep}{6.5mm}
\setlength\bibsep{1pt}

\twocolumn[
  \begin{@twocolumnfalse}
\noindent\LARGE{\textbf{Fluctuation-induced dynamics of multiphase liquid jets with ultra-low interfacial tension}}
\vspace{0.6cm}

\noindent\large{\textbf{Alban Sauret,\textit{$^{ab}$} Constantinos Spandagos,\textit{$^{ac}$} and
Ho Cheung Shum$^{\ast}$\textit{$^{a}$}}}\vspace{0.5cm}

\noindent\textit{\small{\textbf{Received 7th May 2012, Accepted 30th May 2012}}}

\noindent \textbf{\small{DOI: 10.1039/c2lc40524e}}
\vspace{0.6cm}

\noindent \normalsize{Control of fluid dynamics at the micrometer scale is essential to emulsion science and materials design, which is ubiquitous in everyday life and is frequently encountered in industrial applications. Most studies on multiphase flow focus on oil-water systems with substantial interfacial tension. Advances in microfluidics have enabled the study of multiphase flow with more complex dynamics. Here, we show that the evolution of the interface in a jet surrounded by a co-flowing continuous phase with an ultra-low interfacial tension presents new opportunities to the control of flow morphologies. The introduction of a harmonic perturbation to the dispersed phase leads to the formation of interfaces with unique shapes. The periodic structures can be tuned by controlling the fluid flow rates and the input perturbation; this demonstrates the importance of the inertial effects in flow control at ultra-low interfacial tension. Our work provides new insights into microfluidic flows at ultra-low interfacial tension and their potential applications.}
\vspace{0.5cm}
 \end{@twocolumnfalse}
  ]

\section{Introduction}

\footnotetext{\textit{$^{a}$~Department of Mechanical Engineering, University of Hong Kong, Pokfulam Road, Hong Kong \\
E-mail: ashum@hku.hk;
Tel: +852-2859-7904}}
\footnotetext{\textit{$^{b}$~Institut de Recherche sur les Ph\'enom\`enes Hors Equilibre, UMR 7342, \\
CNRS \& Aix-Marseille University, 49 rue F. Joliot Curie, 13013, Marseille, France. }}
\footnotetext{\textit{$^{c}$~Department of Chemical Engineering, Imperial College of Science, \\
 Technology \& Medicine, London, SW72AZ, United Kingdom. }}

Injection of a fluid into a second co-flowing immiscible fluid results in the formation of a jet, which is stable or breaks up into droplets.\cite{1,2} The stability of such jets has been extensively investigated since the earlier work of Plateau\cite{3} and Rayleigh,\cite{4} often motivated by industrial and practical applications such as inkjet printing\cite{5} and emulsion generation.\cite{6} The recent developments of microfluidic technology have enabled the studies of the dynamics and stability of jets at micrometer scales.\cite{7,8} An efficient way to generate such a jet and the resulting monodisperse emulsion relies on the injection of a dispersed phase into a co-flowing continuous phase. In this case, the surrounding fluid has an important impact on the resulting flow regime.\cite{8,9} Two regimes are commonly observed: a dripping regime, where droplets are produced directly after the two fluids meet, and a jetting regime, where droplets are generated after break-up of a jet at some distance downstream. Previous studies of such jets considered Newtonian fluids with a substantial interfacial tension between the two phases, which typically involve a combination of oil-based and water-based liquids. It has led to a good description of the transition between these two regimes; if the interfacial tension effects are sufficiently large compared to viscous and inertial effects, drops are formed.\cite{8,9,10,11,12} However, the actual applications of microfluidics increasingly require the use of fluids with more complex rheological and interfacial properties. Therefore, fundamental studies on multiphase systems with more complicated dynamics are needed; these include fluids with non-Newtonian rheological behaviors\cite{13} and fluids with low interfacial tension.\cite{14}
Among these systems, aqueous two-phase systems (ATPS) typically exhibit an interfacial tension over 500 times lower than in typical oil-water systems. The ATPSs consist of two aqueous phases containing different incompatible additives, such as polymers and polysaccharides; these aqueous mixtures phase-separate at high additive concentrations due to their incompatibility.\cite{15} The presence of the additives leads to two distinct immiscible phases with different properties.\cite{16} The interfacial tension of these systems are very small and they therefore constitute a good example of systems characterized by low interfacial tension. Despite the numerous advantages of all-aqueous systems, such as their biocompatibility and their mild effects to the environment,\cite{15} they remain poorly studied. Indeed, the dynamics of such an interface differ dramatically from classical systems as the interfacial tension effects are several orders of magnitude lower.\cite{17} It prevents the generation of emulsions from a dispersed phase co-flowing with a surrounding continuous phase. The noise inherent to the experimental environment is not sufficient to break up the jets into droplets. To induce droplet formation, we need to actively perturb the dispersed phase; this can be achieved, for instance, with a piezoelectric actuator\cite{14,18} or a mechanical vibrator.\cite{19} Despite the particular emphasis on all-aqueous systems, such control is applicable to any two-phase flows with an ultra-low interfacial tension, including oil-in-water systems in the presence of significant amounts of surfactants.\cite{17}

In this paper, we consider a co-flow jet of two immiscible phases that involve the injection of a dispersed phase into a surrounding continuous phase. We study the dynamics of the jet that forms at the junction where the two fluids meet.8 To date, the characterization of such a system has focused on droplets of oil (or water) in a continuous phase of water (or oil) with interfacial tensions on the order of tens of milli-Newton per meter. In the present study, we consider aqueous two-phase systems (ATPS) as a model system to study the dynamic of the interface. We consider this two-phase flow as it exhibits significantly low interfacial tension. The interfacial tension of different ATPS was measured using a spinning drop tensiometer (Grace Instrument, USA). Based on the measurement, we estimate the interfacial tension of the PEG/K3PO4 system to be lower than $10 \,\mu$N m$^{-1}$. This hampers the formation of emulsion droplets at usual fluid flow rates but leads to new fluidic phenomena, such as the formation of a corrugated jet.
It has been demonstrated experimentally that the transition between dripping and jetting depends on the relative importance of two dimensionless numbers: the Weber number of the inner fluid,

\begin{equation}
We_{in}=\frac{\text{inertial effects}}{\text{interfacial effects}}=\frac{\rho_{in}\,r_{jet}\,{u_{in}}^2}{\gamma}
\end{equation}

and the capillary number of the outer fluid,
\begin{equation}
Ca_{out}=\frac{\text{viscous effects}}{\text{interfacial effects}}=\frac{\eta_{out}\,u_{out}}{\gamma}
\end{equation}
where $r$, $\eta$, $u$, $\gamma$ and $r_{jet}$ are, respectively, density, dynamic viscosity, velocity, interfacial tension and radius of the jet. The subscripts ``in'' and ``out'' denote the dispersed and continuous phases respectively. Dripping occurs for Weber and capillary numbers much smaller than unity.\cite{8} In systems with ultra-low interfacial tensions, the Weber and capillary numbers increase significantly due to the low value of $\gamma$, while the other physical quantities are maintained at the same order of magnitude. In the present study, the interfacial tension is typically 1000 times lower than in classic oil-water interfaces. Thus, the formation of droplets by dripping is not observed and the inertial and viscous effects are more important than interfacial tension effects.

The introduction of a mechanical perturbation of the dispersed phase in a two-phase flow system with an ultra-low interfacial tension leads to two different regimes: generation of droplets and formation of a jet with complicated patterns and corrugations. In the second regime, we observe flow patterns that are unique to systems with ultra-low interfacial tension between the two phases. In this work, we demonstrate that by varying the different operating parameters, we can facilitate the transition from a droplet regime to a corrugated-interface regime. We also characterize the influence of the flow rates of the dispersed and continuous phases as well as the amplitude and frequency of perturbation, achieving a good control over the frequency and the size of the observed corrugations. This method provides an insight into the dynamics of two-phase flow, unique for systems with negligible interfacial tensions.

\section{Experimental methods}

The co-flowing two-phase flow was generated using a glass microcapillary device, which has been described in previous experiments of oil-water emulsion generation,\cite{8} shown schematically in Fig. 1. The inner round capillary, with inner and outer diameters of $200$ $\mu$m and $1.0$ mm (World Precision Instruments, Inc) were tapered to the desired diameter using a micropipette puller (Sutter Instrument, Inc.) to obtain a tip with an approximate inner diameter $d_{tip} \sim 40$ $\mu$m. Then, the tapered round capillary was fitted into a square capillary (Atlantic International Technology, Inc.) with an inner dimension of $1.05$ mm. Matching of the outer diameter of the round capillary to the inner size of the square capillary ensured coaxial alignment of the two capillaries. The dispersed phase was injected into the device through the circular capillary, while the continuous phase was injected in the same direction through the square capillary, leading to a co-flowing jet in the collection capillary. The fluids were injected into the capillaries through flexible plastic tubing (Scientific Commodities Inc.), which were connected to syringe pumps (Longerpump, LSP01-2A) that ensure a controlled flow rate in the microfluidic device. In the present study, the dispersed phase was a 15 wt$\%$ aqueous solution of tripotassium phosphate (K$_3$PO$_4$, Sigma-Aldrich), while the continuous phase was a 17 wt$\%$ aqueous solution of polyethylene glycol (PEG, Sigma-Aldrich). With the specified concentrations, the two phases are immiscible; this is essential for the generation of two-phase flows. In the absence of K$_3$PO$_4$ and PEG, the two fluids directly mix and only form a one-phase flow. The densities of K$_3$PO$_4$ and PEG solutions are, respectively $1159$ kg m$^{-3}$ and $1027$ kg m$^{-3}$, while their dynamic viscosities are $1.46$ mPa s$^{-1}$ and 16.2 mPa s$^{-1}$ respectively. The interfacial tension between the two aqueous phases is very low, around $\gamma \sim 10 \, \mu$N m$^{-1}$.

\begin{figure}
\begin{center}
\includegraphics[width=8.5cm]{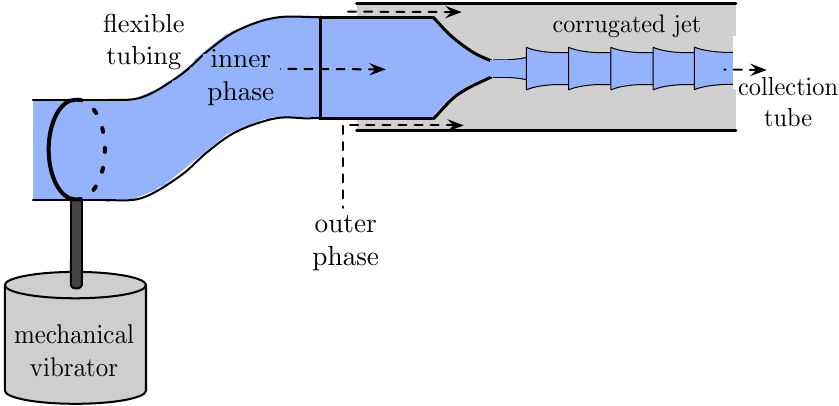}
 \caption{Schematic of the experimental set-up used in this study. The flexible tubing where the dispersed phase is introduced into the cylindrical capillary in the device is connected to a mechanical vibrator, which provides an external forcing into the system. In the absence of vibrations, a jet is formed when the dispersed and continuous phases meet inside the square capillary. When pressure perturbations are applied, droplets or a corrugated jet are observed.}
\label{setup}
\end{center}
\end{figure}

The plastic tubing, which introduces the dispersed phase in the device, was connected to a mechanical vibrator (Pasco Scientific, Model SF-9324). The vibrator shook the flexible tubing; this induces pressure fluctuation to the dispersed phase. A function waveform generator (Rigol, DG1012) connected with a power amplifier drove the vibrator to induce different signals with tunable amplitudes and frequencies (from $0.1$ Hz to $50$ kHz). In the present study, a sinusoidal signal was used in the range [$1-50$] Hz. Vibration of the tip of the inner capillary connected to the plastic tubing was not observed, ensuring that the effect of the mechanical vibrator was limited only to varying the pressure of the dispersed phase.
The flow behavior inside the microcapillary device was monitored with an inverted microscope (Motic, ocular: WF10 $\times$ 18 mm, object lens: EA4). A high-speed camera (Photron FASTCAM SA4) was connected to the microscope and captured the flow through the capillary. For some experiments, we added methylene blue to the inner fluid for enhanced flow visualization. We did not see any significant difference in the fluid dynamic between the experiments performed with and without the methylene blue.

\section{Results}

\subsection{From drops to corrugations}

We show that the dynamics of the interface at ultra-low interfacial tension is modified by the application of an external perturbation at controlled frequencies. In the absence of vibrations, and with typical flow rates of the dispersed, $Q_{in}$, and continuous, $Q_{out}$, phases, for instance, $Q_{in} = 150$ $\mu$L h$^{-1}$ and $Q_{out} = 5000$ $\mu$L h$^{-1}$, a stable jet of radius, $r_{jet}$, is observed in the junction where the two fluids meet. Due to the very low interfacial tension, no growth of disturbance is observed all along the jet. This is in agreement with a very low value for the growth rate associated to a Rayleigh-Plateau instability (see Fig. 2a). As we introduce vibrations into the system, at low frequencies of f = 3 Hz, the interface between the two fluids oscillates. The oscillation has a well-defined wavelength, which is directly related to the frequency of vibration of the tubing, but the jet does not break up into droplets (Fig. 2b). The wavelength of the perturbation, the flow rates of the inner fluid and the applied frequency satisfy the following relation:

\begin{equation}
\lambda=\frac{Q_{in}}{\pi\,{r_{jet}}^2\,f} \simeq 950\,\mu\text{m}
\end{equation}

This theoretical wavelength agrees well with the measured wavelength of approximately $940$ $\mu$m (see Fig. 2b). As we increase the frequency further to $f = 6$ Hz, monodisperse droplets are generated and subsequently collected.\cite{20} The interface between the droplets is well defined, but the shape of the droplets is not perfectly spherical. As the inertial force is large with respect to the interfacial tension, capillary force is not sufficient to restore the spherical shape of the droplets (Fig. 2c). An even further increase in the frequency to $f = 9$ Hz results in the formation of a corrugated interface (Fig. 2d). These corrugations resemble morphologically the ones observed in water-lubricated transport of oil inside horizontal pipes\cite{21} and in droplet-induced corrugations in a three-phase core-annular flow.\cite{17} The shape of the interface evolves along the collection tube as the corrugations are advected by the Poiseuille flow. Subsequently, at an even larger frequency of $f = 15$ Hz, we observe that the frequency of the corrugations also increases while their amplitude becomes smaller (Fig. 2e).

\begin{center}\begin{figure}[h!]
\includegraphics[width=8cm]{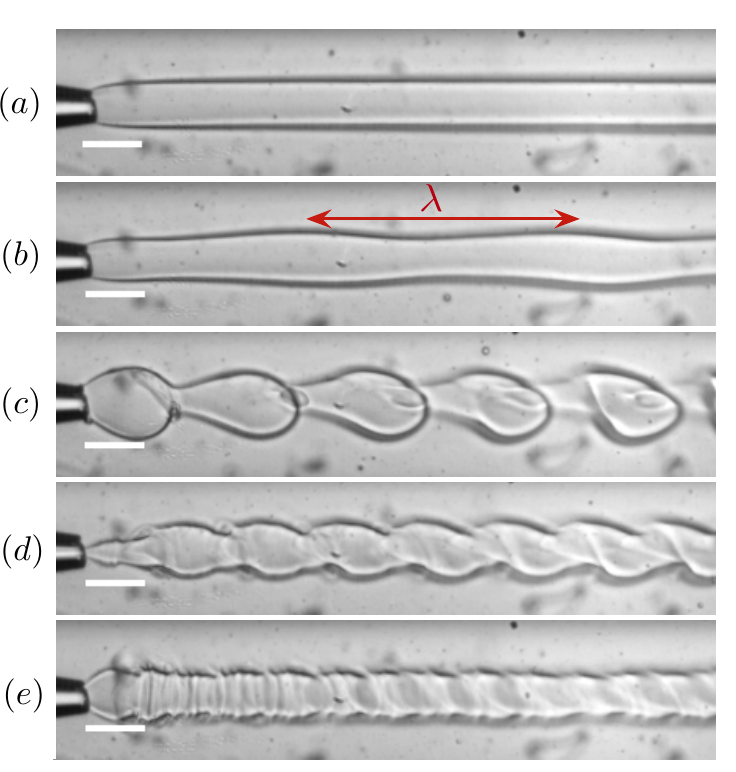}
 \caption{Different flow regimes at increasing frequency of applied vibration to the dispersed phase. Flow rates are $Q_{in} = 150\,\mu$L h${-1}$ and $Q_{in} = 5\,000\,\mu$L h${-1}$: (a) At $f = 0$ Hz, a stable jet is formed. (b) At $f = 3$ Hz, we observe oscillations of the jet. (c) At a higher frequency of $f = 6$ Hz, monodisperse droplets are generated. (d) At an even higher frequency of $f = 9$ Hz, corrugations are formed. (e) At a frequency of $f = 15$ Hz, corrugations with a smaller wavelength are observed. The dispersed phase is 15 wt$\%$ K$_3$PO$_4$ in water, the continuous phase is 17 wt$\%$ PEG in water. The fluids flow from left to right. Scale bars are $200\,\mu$m.}
\end{figure}\end{center}

\subsection{Transition between drop formation and corrugated jet}

To capture the physical mechanism underlying the dependence of the morphology of the jet on the disturbance frequency, we performed a systematic experimental study varying both the amplitude and the frequency of the pressure disturbance, keeping the inner ($Q_{in}$) and outer ($Q_{out}$) flow rates constant while maintaining a low $Q_{in}/Q_{out}$ ratio: $Q_{in} = 150$ $mu$L h$^{-1}$ and $Q_{out} = 5000$ $mu$L h$^{-1}$. As we shall see later, the ratio $Q_{in}/Q_{out}$ has an important influence of the observed flow regime. Here, two main flow regimes can be distinguished (see Fig. 3): a regime where drops are generated and a corrugated jet regime. For the present fluid flow rates, the optimal frequency for forming droplets lies in the range $f = 4$ Hz and $f = 5$ Hz where droplets are formed at a wide range of driving amplitude. At frequencies larger than $8$ Hz, the driving amplitude achievable with our current experimental setup is not sufficient to break the jet into droplets. The amplitude of pressure disturbance is directly related to the amplitude of the shaking of the inner tubing, which is tuned by varying the input voltage of the vibrator. However, a direct relationship between the amplitude of shaking of an oscillating flexible tubing and the resulting pressure perturbation remains challenging to obtain.\cite{22} The occurrence of these different regimes is presented in a ($U,f$) state diagram where $U$ is the driving amplitude of the mechanical vibrator in volts and $f$ the frequency in Hz, in Fig. 3.
An estimation of the optimal frequency required to break to inner jet into droplets can be done based on the Rayleigh- Plateau instability analysis. For an inviscid jet in the air, a theory has been developed by Lord Rayleigh,\cite{2,4} who considered a circular liquid jet. This jet can become unstable due to the capillary force and break up into droplets. This simple case already leads to a good estimate of the optimal frequency to break the jet into droplets, which corresponds to the maximum value of the growth rate for the Rayleigh-Plateau instability. We considered the flow in the jet as incompressible and introduced a perturbation $v = v_0+\delta\,v$ and $p = p_0+\delta p$, where $v_0$ is the mean velocity of the fluid and $p_0$ is the Laplace pressure in the undisturbed jet $p_0 = \gamma/r_{jet}$ with $\gamma$ being the interfacial tension. The radius of the jet in the frame of reference at the velocity, $v_0$, writes $r_{jet}(t) = r_0 (1 + \epsilon_0\,\text{e}^{\sigma\,t})$, where $r_0$ is the radius of the undisturbed jet. A linear stability analysis enables us to calculate the growth rate of the capillary instability:\cite{2,4}
\begin{equation}
\sigma_{NV}=\left[\frac{\gamma\,k}{\rho\,{r_{jet}}^2}\left(1-k^2\,{r_{jet}}^2\right)\,\frac{\text{I}_1(k\,r_{jet})}{\text{I}_0(k\,r_{jet})}\right]^{1/2}
\end{equation}

\begin{center}\begin{figure}
\includegraphics[width=8.5cm]{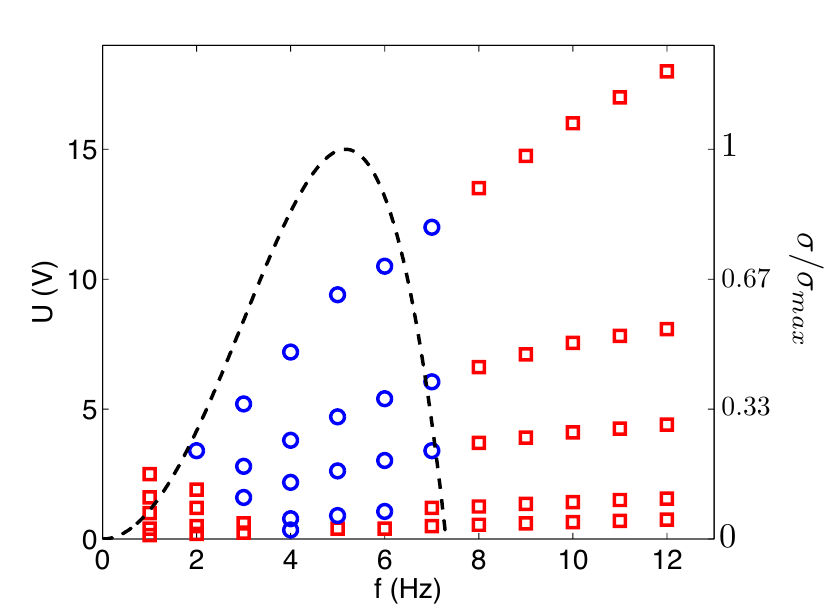}
 \caption{Phase diagram of the different flow regimes encountered in the experiments as a function of the driving amplitude $U$ (in volts) and the frequency $f$ (in Hz) for an inner and outer flow rates of $Q_{in} = 150\,\mu$L h${-1}$ and $Q_{in} = 5\,000\,\mu$L h${-1}$ respectively. Red squares indicate the regime of corrugated jets, blue circles are the regime where monodisperse drops are formed. The black dashed-line is the rescaled growth rate of the Rayleigh-Plateau instability $\sigma/\sigma_{max}$ obtained using the theoretical analysis of Guillot et al.\cite{9,12}.}
\end{figure}\end{center}

where $r_0$ is the density of the fluid, $k$ is the wavenumber, I$_0$ and I$_1$ are, respectively, the Bessel functions of order 0 and 1. The wavenumber $k$ and the frequency of perturbation $f$ satisfy the relation $k = 2\,\pi\,f/v_{jet}$, where $v_{jet}$ is the velocity of the inner fluid determined by the applied fluid flow rates. From this expression of the growth rate, the maximum of the function $\sigma(k)$ is reached for $k\,r_{jet} \sim 0.7$; thus the most amplified wavenumbers is $k_m\,r_{jet} \sim 0.7 r_{jet}$ whereas for $k\,r_{jet} > 1$, no perturbation can grow as the growth rate, $\sigma(k)$, is imaginary. Later, Tomotika\cite{23} has shown that the presence of a surrounding fluid has an important effect on the stability of the cylindrical thread. However, in this case, the outer fluid is at rest and therefore does not describe the influence of the velocity of the outer fluid. Therefore, we have considered in this paper the analysis of Guillot et al.\cite{9,12,24} who considered a co-flowing jet confined in a microchannel similar to our system. Using the theory of a confined jet allow us to directly obtain the growth rate of the Rayleigh-Plateau instability for a given radius of the outer channel, fixed inner and outer fluid flow rates, and fluid properties. In this case, the growth rate of the Rayleigh-Plateau instability $\sigma$ is given by
\begin{equation}
\sigma=\frac{\gamma}{16\,\eta_{out}\,R}\,\frac{F(x,\lambda)\,(k^2-k^4)}{x^9\,(1-\lambda^{-1})-x^5}
\end{equation}
with
\begin{eqnarray}
F(x,\lambda)=x^4\,(4-\lambda^{-1}+4\,\text{ln}x)+x^6\,(4\,\lambda^{-1}-8) \nonumber \\
+x^8\,\left[4-3\,\lambda^{-1}-4\,(1-\lambda^{-1})\,\ln x\right]
\end{eqnarray}
where $\lambda = \eta_{in}/\eta_{out}$ as the ratio of the dynamic viscosities of the inner and outer fluids, respectively. $R$ is the radius of the cylindrical outer capillary, $k$ is the dimensionless wavenumber of the perturbation, and $x$ represents the dimensionless inner thread radius $r_{jet}/R$. This growth rate of the capillary instability is calculated by applying measured values of the different variables to equations (5) and (6). The resultant growth rate is plotted as a function of the frequency of perturbation $f = 2\,\pi\,v_{jet}/k$ in Fig. 3. The mechanism responsible for the break up of the inner thread remains the same in all three approaches; the only difference is that in the approach by Guillot et al.,\cite{9,12} the value of the growth rate is modified by the presence of an outer phase with a non-zero velocity.
Apart from the frequency and amplitude of the driving fluctuation, the ratio of flow rates of the two phases also affects the morphology of the jets. For all flow rates, we tune the frequency and the amplitude of pressure disturbance to determine whether drop formation is possible. For a given outer flow rate $Q_{out}$, formation of drops is achieved at sufficiently low flow rates of the inner fluids; above the critical flow rate, droplets cannot be generated and instead, a corrugated jet is formed. This can be explained by the resulting thickness of the jet.
As the ratio $Q_{in}/Q_{out}$ increases, the thickness of the jet increases accordingly.\cite{8} Therefore, since the interfacial tension effects are negligible, the deformation of the interface must be large enough to chop off the jet into droplets; otherwise, no droplets can be formed. The limit of the degree of deformation achievable is imposed by the experimental setup. The results are summarized in a phase diagram, which shows the different flow regimes as a function of $Q_{in}$ and $Q_{out}$, in Fig. 4. In the following characterization of corrugated jets, we impose a ratio of the inner fluid flow rate to the outer fluid flow rate ($Q_{in}/Q_{out}$) such that no drop formation is observed. This allows us to generate a corrugated jet and study the influence of the different relevant parameters on its dynamics.

\begin{center}\begin{figure}
\includegraphics[width=8.5cm]{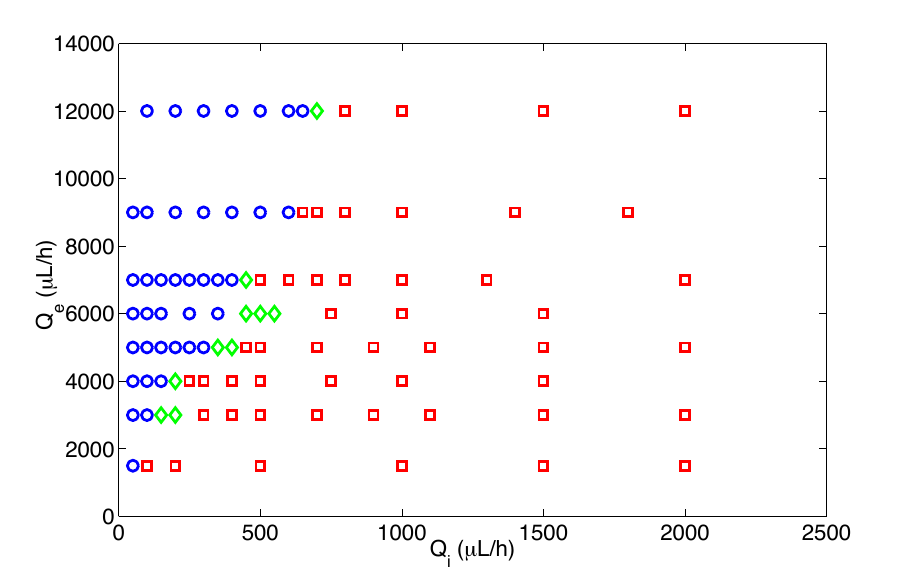}
 \caption{Phase diagram of the dynamics of the co-flow jet as a function of the inner ($Q_{in}$) and outer ($Q_{out}$) flow rates. Blue circles refer to the regime where monodisperse drops can be formed whereas the red squares indicate the regime where no drop formation is possible for any frequency and amplitude of perturbation leading to a corrugated jet.}
\end{figure}\end{center}

\subsection{Tuning the shape of corrugations}

In the corrugated-jet regime, the shape of the corrugations can be manipulated by adjusting the fluid flow rates as well as the frequency of perturbations. The fluid-fluid interface becomes highly corrugated if the inner fluid flow rate, $Q_{in}$, is significantly increased, for example, to $1000$ $\mu$L h$^{-1}$ at a constant outer fluid flow rate, $Q_{out} = 5000$ $\mu$L h$^{-1}$, as shown in Fig. 5a. For given inner and outer fluid flow rates, an increase in the frequency of perturbation leads to a corresponding increase in the number of corrugations per unit length (Fig. 5b-d). The observed frequency of the corrugations is directly related to the applied frequency (see Fig. 5e). Moreover, the amplitude of the applied perturbation can also be used to change the amplitude of the corrugations, as exhibited in Fig. 6. An increase in the perturbation amplitude leads to corrugations with higher amplitudes.

\begin{center}\begin{figure}[h!]
\includegraphics[width=8cm]{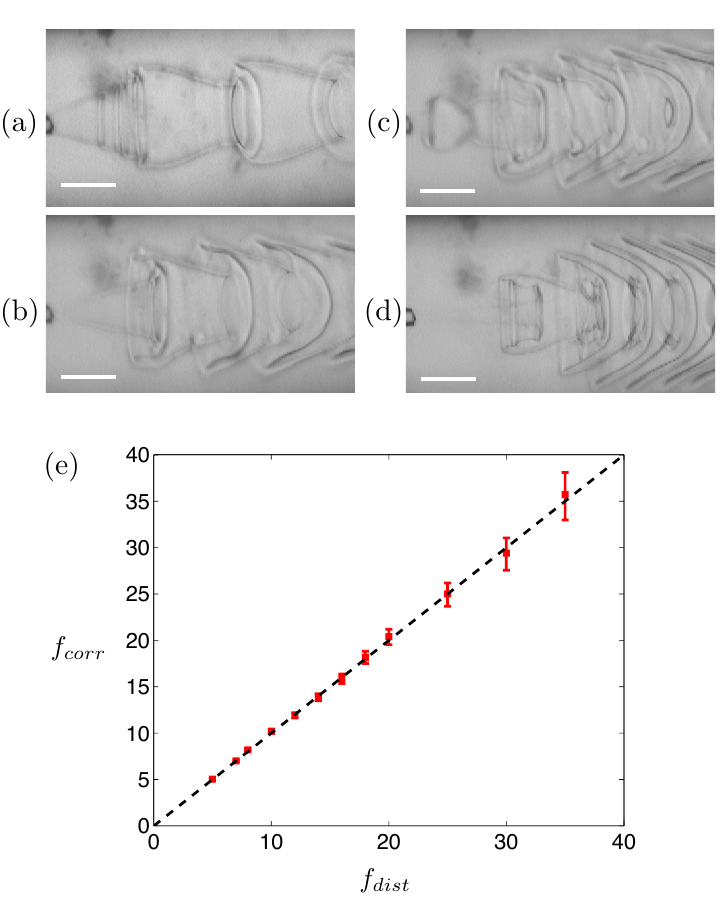}
 \caption{The interface between the two fluids becomes highly corrugated at high flow rates and no droplet formation is observed: $Q_{in} = 1\,000\,\mu$L h${-1}$ and $Q_{in} = 5\,000\,\mu$L h${-1}$. Here $Q_{in}$ is seven times larger than that in Fig. 2. The number of corrugations increases with increasing frequency. (a) $f = 7$ Hz, (b) $f = 9$ Hz, (c) $f = 11$ Hz, (d) $f = 20$ Hz. The dispersed phase is 15 wt$\%$ K3PO4 in water, while the continuous phase is 17 wt$\%$ PEG in water. Scale bars are $200\,\mu$m. (e) Frequency of corrugations observed, $f_{corr}$, as a function of the applied disturbance frequency, $f_{dist}$. The black dashed line has a slope of unity.}
\end{figure}\end{center}

\begin{center}\begin{figure}[h!]
\includegraphics[width=8cm]{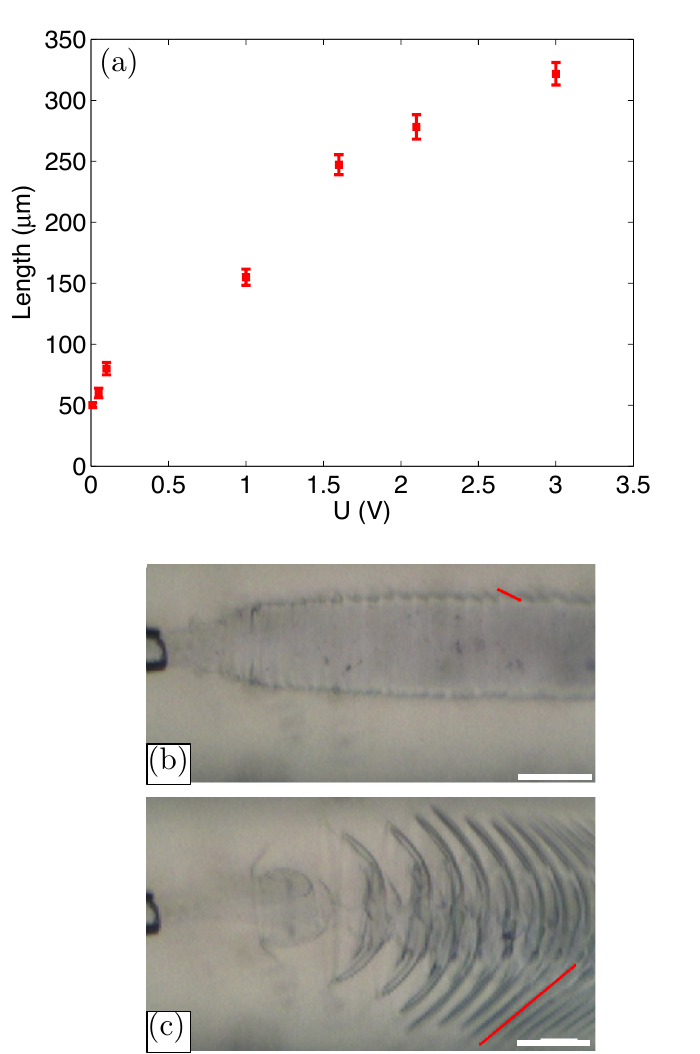}
 \caption{(a) The effect of increasing the driving amplitude of mechanical vibrator on the size of the corrugations, for given frequencies f = 11 Hz and flow rates $Q_{in} = 1\,000\,\mu$L h${-1}$ and $Q_{in} = 5\,000\,\mu$L h${-1}$. An increase in the voltage from (b) 0.1 V to (c) 2.2 V, results in thinner and longer corrugation ÔÔtailsÕÕ. The dispersed phase is 15 wt$\%$ K3PO4 in water, while the continuous phase is 17 wt$\%$ PEG in water. Scale bars are $200\,\mu$m.}
\end{figure}\end{center}

\subsection{Discussions}
The transition between drop formation and corrugated jet formation can be understood in terms of a Rayleigh-Plateau instability. We calculate the theoretical growth rate for a co-flowing jet; its variation with respect to the frequency is given in Fig. 3. This simple theory predicts that no drops formation can be observed for frequency $f \geq 8$ Hz with the considered flow rates because the excited wavelength is such that $\lambda \geq 2\pi/{r_{jet}}^2$ This is in agreement with the experimental results shown in Fig. 2. For low frequency, the growth rate remains small and thus the time required to break the jet becomes longer. This explains why droplets are not produced experimentally, for frequency $f \leq 2$ Hz. For frequency around $f = 4-5$ Hz, the growth rate is maximum and droplets are formed. The value of the optimal frequency for which the growth rate is maximum can be roughly estimated using the Rayleigh-Plateau criterion for an inviscid jet in the air: $k_{rjet} \simeq 0.7$, which leads to the relation for the perturbation frequency:
\begin{equation}
f \simeq \frac{0.7\,Q_{in}}{2\,\pi\,{r_{jet}}^2}
\end{equation}

According to this systematic study, the most effective frequency for inducing droplet formation is the optimal frequency, $f \simeq 4.5$ Hz. It shows a significant deviation with the estimation based on the Rayleigh-Plateau instability for an inviscid jet in air. To obtain a better value for this optimal frequency, we need to consider the influence of the surrounding fluid with a non-zero-velocity. Unfortunately, there is no simple expression in this case.\cite{9,12} The theoretical expression of the growth rate in this case is shown in Fig. 3 and exhibits a good agreement with the experimental results. This confirms that the mechanism responsible for the break up of the perturbed jet in our experiments is Rayleigh-Plateau instability. Unlike in the undisturbed case where no perturbation would grow, disturbance of the jet can be induced by introducing an initial disturbance at appropriate amplitudes.
However, when the size of the jet becomes too large or when the applied frequency is not optimal, no droplet formation can occur. In this case, the pressure perturbation initially modulates the shape of the interface and the driving force behind the further deformation of the interface and the subsequent formation of the corrugations is the pressure fluctuation applied to the system via the mechanical vibrator. The interfacial tension between the dispersed and the continuous phase is very low, and thus inertia dominates. Therefore, the interface between these two aqueous phases is highly deformable, and is susceptible to external forcing. The mechanism behind this corrugated jet is as follows: when the pressure of the dispersed phase reaches its minimum, the jet exiting the nozzle has a small diameter (Fig. 7a). As the driving pressure increases, the flow rate of the dispersed phase increases and so does the jet diameter (Fig. 7b). In this manner, the diameter of the jet changes periodically according to the applied perturbation. Thus the wavelength of the oscillations of the jet follows that of the applied perturbation. The deformed interface is then advected by the Poiseuille flow in the microchannel (Fig. 7c) and continues to deform passively due to the ultra-low surface tension between the two aqueous phases. Downstream in the collection tube, the corrugations appear as sharp edges (Fig. 7d). To sum up, the applied pressure is proportional to the velocity of the inner jet due to the low Reynolds number of the flow (typically around unity). As a result, the observed corrugations are not due to any novel source of instability. In this way, the jet adopts the shape of the corrugations that resemble the shapes of ÔÔarrowsÕÕ and ÔÔmushroomsÕÕ. The mechanism behind the observed phenomenon of the formation of corrugations in water-water interfaces, is similar to the one proposed previously for the formation of corrugations in a three-phase core-annular flow.\cite{17} In that case, the moving inner droplets deform the jet interface and the subsequent advection of the interface by a Poiseuille flow induces the corrugations. Here, the same mechanism is at play; instead of the periodic generation of the inner droplets, the pressure fluctuation introduces the initial deformation of the interface. References Thus the frequency of corrugations exactly matches that of the applied pressure disturbance, as confirmed in Fig. 5e.

\begin{figure}[h!]\begin{center}
\includegraphics[width=7cm]{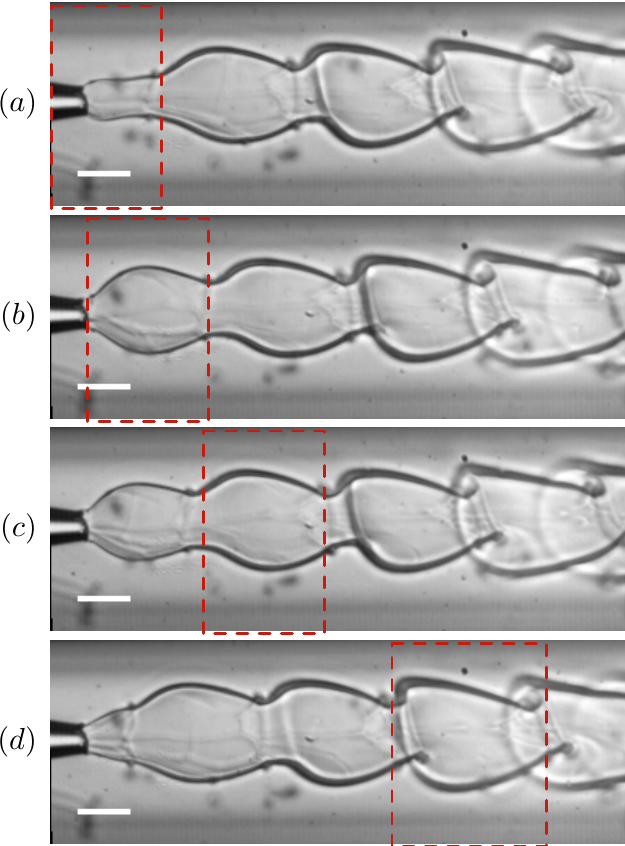}
 \caption{Evolution of the corrugated jet for an inner and outer flow rates of $Q_{in} = 600\,\mu$L h${-1}$ and $Q_{in} = 5\,000\,\mu$L h${-1}$ respectively. The frequency used is f = 5 Hz and the driving amplitude $U = 1.4$ V. The red boxes follow a specific corrugation. Scale bars are $200\,\mu$m.}\end{center}
\end{figure}

Depending on the flow rate, the amplitude and the frequency of perturbation, different corrugation shapes are obtained, as presented in Fig. 8. The control over the morphology of the corrugations creates new opportunities to generate complex materials such as fibers with an increased surface area with respect to their volume for applications related to materials engineering and catalysis.

\section{Conclusions}

In this work, we characterize the drop-to-corrugated jet transition with respect to the driving frequency and amplitude as well as the fluid flow rates in systems where the interfacial tension between the two immiscible phases is very low (lower than $10 \mu\,$N m$^{-1}$). Previously, similar corrugations have been formed by introducing inner droplets inside a jet.\cite{17} Since it is not always easy to introduce inner droplets in a controlled fashion, our approach offers a more versatile route to perturb a jet. Moreover, the present approach, which relies on a perturbation of the inner phase, is simple and enables a high degree of control over the flow rates and the frequency and amplitude of the external forcing. Adjustment of these parameters allows the manipulation of the fluidic regime. We also observe new corrugation morphologies characterized by thinner and longer tails, different from those reported previously. This new understanding in the perturbation of liquid jets inspires new strategies to manipulate liquid systems with ultra-low interfacial tensions and leads to the development of methods for creating novel fibers and droplets that can be used in various industrial and biomedical applications. Applying this method to fluids that can be instantaneously solidified would generate new material structures at the micrometer scale in a controlled fashion.

\section*{Acknowledgements}

This research was supported by the Seed Funding Programme for Basic Research (201101159009) and Small Project Funding (201109176165) from the University of Hong Kong. We thank L. Xu for the use of a high speed camera in some experiments.

\bibliography{rsc} 
\bibliographystyle{rsc} 

\begin{figure*}\begin{center}
\includegraphics[width=14cm]{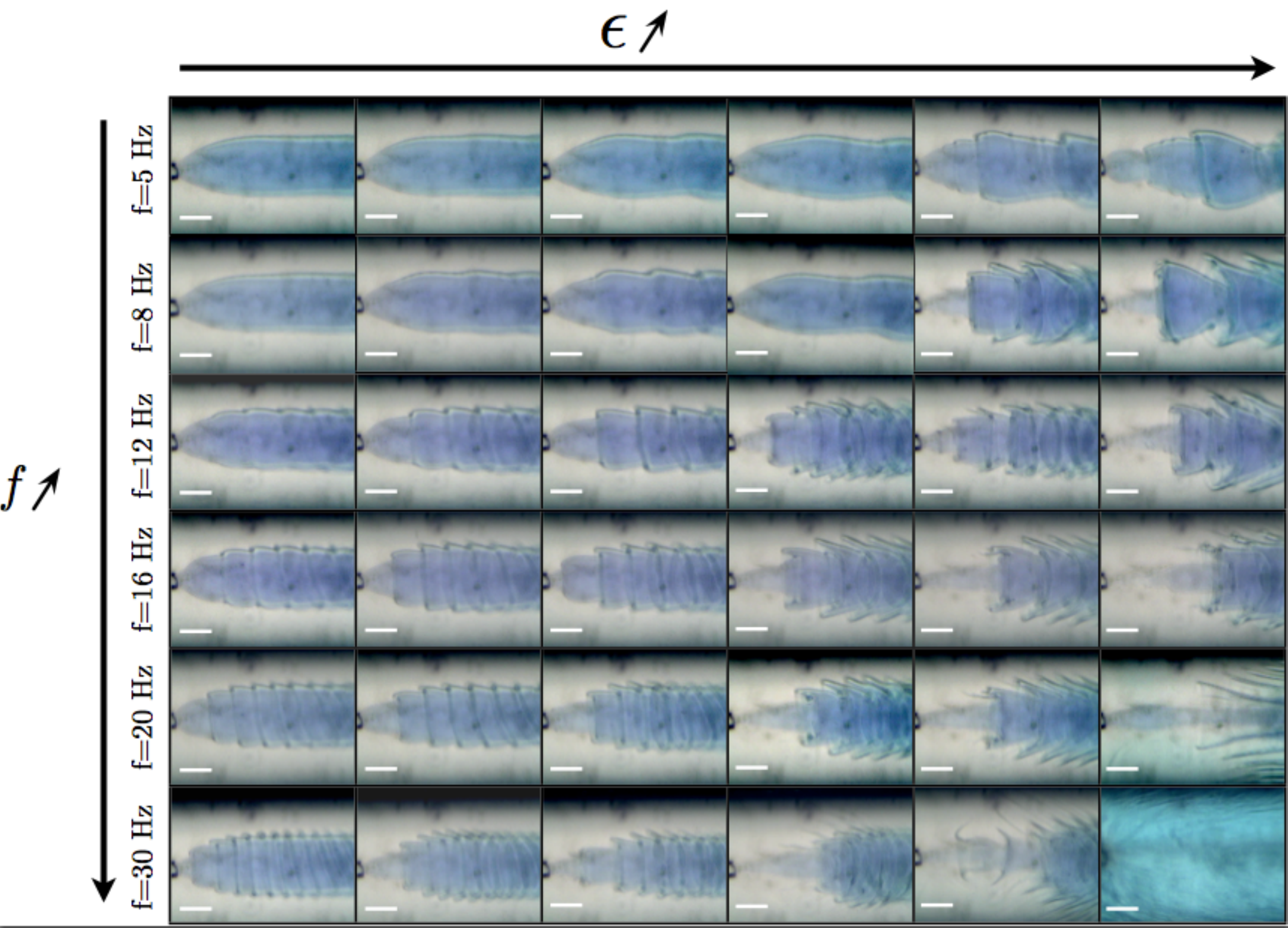}
 \caption{State diagram of corrugated jets as a function of perturbation frequency and amplitude for inner and outer flow rates of $Q_{in} = 2\,000\,\mu$L h${-1}$ and $Q_{in} = 5\,000\,\mu$L h${-1}$ respectively. }
\end{center}\end{figure*}

\end{document}